\documentclass[epj]{svjour}
\RequirePackage{graphicx}
\RequirePackage{epstopdf}
\RequirePackage{amssymb}
\RequirePackage{bbm}
\RequirePackage{slashed}
\RequirePackage{float}
\RequirePackage{subfigure}
\RequirePackage{amsmath}
\RequirePackage[colorlinks,citecolor=blue,urlcolor=blue,linkcolor=blue]{hyperref}
\RequirePackage{amssymb}
\hypersetup{pdfstartview=}
\begin{document}

\title{Bounds on the absorptive parts of the  chromomagnetic and chromoelectric dipole moments of the top quark from LHC data}
\author{A. I. Hern\'andez-Ju\'arez    \inst{1}\thanks{alaban7\_3@hotmail.com}\and
        A. Moyotl  \inst{2}\thanks{agustin.moyotl@uppuebla.edu.mx}\and
        G. Tavares-Velasco\inst{1}\thanks{gtv@fcfm.buap.mx}
}

\institute{Facultad de Ciencias F\'isico-Matem\'aticas,\\
  Benem\'erita Universidad Aut\'onoma de Puebla,\\
 C.P. 72570, Puebla, Pue., Mexico \label{addr1} \and Ingenier\'ia en Mecatr\'onica,\\ Universidad Polit\'ecnica de Puebla,\\ Tercer Carril del Ejido Serrano s/n, San Mateo Cuanal\'a, Juan C. Bonilla,\\ Puebla, Puebla, M\'exico \label{addr2}}
 \offprints{A. I. Hern\'andez-Ju\'arez (corresponding author)}
\mail{alaban7\_3@hotmail.com}
\date{Received: date / Accepted: date}

\abstract{
Bounds on the  absorptive (imaginary) parts of the top quark chromomagnetic $\hat{\mu}_t$ and chromoelectric $\hat{d}_t$ dipole moments are obtained by reinterpreting  the most recent LHC data in top quark pair production. It is  found that both limits  are of the  order  of $10^{-1}-10^{-2}$, which are consistent with the standard model  prediction of ${\rm Im}\big[\hat{\mu}_t\big]$. The effects of the absorptive parts of the top quark dipole moments  are also studied via some kinematic distributions of  $\overline{t}t$ production, though no significant deviation from the standard model  leading order contribution is observed. Our bounds can be useful to constrain the parameter space of standard model extensions. 
}



\maketitle

\section{Introduction}
Quite recently, the study of the chromomagnetic dipole moment (CMDM) $\hat{\mu}_t$ of the top quark has become a topic of great interest both theoretically and experimentally. On the theoretical side, a new evaluation of the lowest order contributions to $\hat \mu_t$ within the framework of the standard model (SM) was presented in Refs. \cite{Hernandez-Juarez:2020drn,Aranda:2020tox}, which has  settled some ambiguities found in previous evaluations:  in contrast  to what  was claimed before \cite{Martinez:2007qf}, it has become  clear that the CMDM is infrared divergent, with the divergent part arising from the non-abelian term of the gluon field  tensor \cite{Hernandez-Juarez:2020drn,Aranda:2020tox}. Therefore, the study of the static CMDM has  no sense in perturbative QCD. Nevertheless,  the  off-shell CMDM is finite and gauge independent in the SM \cite{Hernandez-Juarez:2020drn}, and therefore it can be a valid observable quantity. In addition  several non SM contributions to the CMDM have been calculated up to the one-loop level in the framework of extension theories such as  two Higgs doublet models (THDM) \cite{Gaitan:2015aia},  fourth-generation THDMs \cite{Hernandez-Juarez:2018uow}, 331 models \cite{Hernandez-Juarez:2020xon},  models with an extra $Z$ gauge boson \cite{Aranda:2018zis}, etc.  
As far as the top quark chromoelectric dipole moment (CEDM) $\hat{d}_t$ is concerned, it is induced up to the three-loop level in the SM \cite{Czarnecki:1997bu} and thus could give a clear signal of $CP$ violation. The top quark CEDM has also been a topic of  interest the literature as it can arise at  one-loop level in some beyond the SM (BSM) theories \cite{Hernandez-Juarez:2018uow,Hernandez-Juarez:2020xon}, thereby opening the possibility of a considerable enhancement. In general, both the off-shell CMDM and CEMD can have non-zero imaginary parts, whose effects  remain almost unexplored.
 
On the experimental side,  the leading order corrections to the cross section of top quark pair production induced by the top quark CMDM and CEDM have been studied in \cite{Atwood:1994vm,Simmons:1995hb,Haberl:1995ek,Cheung:1995nt,Choi:1997ie,Hikasa:1998wx,Czakon:2008ii,Antipin:2008zx,Gupta:2009eq,Zhang:2010dr,Degrande:2010kt,Baumgart:2012ay,Englert:2012by,Hayreter:2013kba,Bernreuther:2013aga,Fabbrichesi:2014wva,Aguilar-Saavedra:2014iga,Bernreuther:2015yna,Cao:2015doa,Barducci:2017ddn,Malekhosseini:2018fgp}, and the next-to-leading order corrections have also been calculated more recently \cite{Englert:2014oea,Franzosi:2015osa,AguilarSaavedra:2018nen}, whereas the effects of the dipole moments have been analyzed in some processes \cite{Kane:1991bg,Atwood:1992vj,Grzadkowski:1997yi,Lampe:1997sj,Yang:1997iv,Tsuno:2005qb,Hioki:2009hm,Gupta:2009wu,Hioki:2010zu,HIOKI:2011xx,Kamenik:2011dk,Hioki:2012vn,Hioki:2013hva,MammenAbraham:2021ssc}. The CMS collaboration has imposed the following current bounds on  the top quark CMDM and CEDM:  $-0.014<\hat{\mu}_t<0.004$ and $-0.020<\hat{d}_t<0.012$ \cite{CMS:2018jcg}, which where obtained via two opposite sign leptons ($e^+e^-$, $e^{\pm}\mu^\mp$, $\mu^+\mu^-$) in the final state. Furthermore,  the CMS collaboration  also set the limits  $\hat{\mu}_t=-0.024^{+0.013}_{-0.009}\text{(stat)}^{+0.016}_{-0.011}\text{(syst)}$ and $|\hat{d}_t|<0.03$ \cite{Sirunyan:2019eyu} obtained by the analysis of lepton+jets events in the final state.  These bounds were extracted from experimental data by assuming that the  top quark CMDM and CEDM are real quantities. 

An appropriate approach to study the anomalous  couplings and their effects on observable processes in a model-independent way is provided by the effective Lagrangian approach, where a $SU(3)_c\times SU(2)_L\times U(1)_Y$ gauge-invariant effective Lagrangian is introduced to  parametrize the effects of physics BSM. Such an effective Lagrangian contains the SM Lagrangian plus  a tower of  effective operators $\mathcal{O}^n$ ($n> 4$) constructed out of the SM fields
\begin{equation}
\mathcal{L}^{\rm Eff.}=\mathcal{L}^{\text SM}+\sum_{n>0}\frac{\alpha_\mathcal{O}}{\Lambda^{n+4}}\mathcal{O}^{n+4}, 
\end{equation}
where the coupling constants $\alpha_\mathcal{O}$ parametrizes our ignorance of the new physics and $\Lambda$ is the new physics scale. In particular,  non-standard top-gluon interactions arise from a dimension-six operator \cite{Arzt:1994gp,Aguilar-Saavedra:2009ygx}, which after electroweak symmetry breaking gives rise to the the following Lagrangian  
\begin{equation}
\label{ttgInteraction}
\mathcal{L}=-g_s\bar{t}T^a\Big[\frac{\sigma^{\mu\nu}}{2m_t}\Big(\hat{\mu}_t+i\hat{d}_t  \gamma^5\Big)G^a_{\mu\nu}\Big]t,
\end{equation}
where $T^a$ are the $SU(3)$ color generators, $G^a_\mu$ are the gluon fields, $G^a_{\mu\nu}=\partial_\mu G^a_{\nu}-\partial_\nu G^a_{\mu}-g_s f_{abc}G^b_\mu G^c_\nu$ is the gluon field tensor, whereas $\hat{\mu}_t$ and $\hat{d}_t$ are constant coefficients that parametrizes the anomalous contributions to the $ttg$ coupling arising from new physics.
It must be noted however that the  above Lagrangian does not yield the most general $\overline{t}tg$ interaction, which in fact can be written in terms of four independent form factors \cite{Nowakowski:2004cv}. 

The above Lagrangian also describes the interaction between an off-shell gluon and two on-shell quarks \cite{Davydychev:2000rt}. Since the off-shell CMDM and CEDM of the top quark are complex in general, they will be written as
\begin{align}
\hat{\mu}_t&={\rm Re}\big[\hat{\mu}_t\big]+i {\rm Im}\big[\hat{\mu}_t\big],\\ \hat{d}_t&={\rm Re}\big[\hat{d}_t\big]+i  {\rm Im}\big[\hat{d}_t\big]. 
\end{align}
As far as  the SM predictions are concerned, the real and imaginary parts of the off-shell CMDM of the top quark are of the order of $10^{-2}-10^{-3}$ \cite{Hernandez-Juarez:2020drn}, whereas the predictions for the off-shell CEDM are not available yet. Nevertheless, in  BSM theories both real and imaginary parts of the off-shell top quark CEDM are of the order of $10^{-19}$ \cite{Hernandez-Juarez:2020xon}. On the other hand, the effects of the absorptive parts of the CMDM and CEDM at LHC  were first studied in \cite{Bernreuther:2013aga} but to our knowledge, there is no update on such analysis, which we believe is in the order given the current experimental bounds on these observables.

We  would like to note that although the off-shell  dipole form factors are dependent on the gluon transfer momentum, we will follow the authors of Ref. \cite{Bernreuther:2013aga} and consider in our analysis below that both  the real and absorptive parts of the dipole form factors are constant, which is valid as long as unitarity is not spoiled.  This approach has also been used for instance to obtain bounds on the trilinear neutral gauge boson couplings $Z\gamma V^*$ ($V=\gamma, Z$) \cite{Baur:1992cd}, which due to Bose statistics and angular momentum conservation are non-vanishing only for off-shell $V$. In fact,  over a large interval of $q^2$, the SM contribution to the top quark CMDM shows little variation \cite{Hernandez-Juarez:2020drn}, and the same is true for both the CMDM and CEDM in some BSM theories (see for instance \cite{Hernandez-Juarez:2020xon,Aranda:2018zis}).   Our main goal is to obtain bounds on  ${\rm Im}\left(\hat{\mu}_t\right)$ and $ {\rm Im}\left(\hat{d}_t\right)$ using the  data  for top quark pair production at the LHC run 2, which in turn can be useful to constraint the parameter space of some BSM theories.

  Our work is organized as follows. In Sec. \ref{SecComments} we discuss the framework for the study of the CMDM and CEDM absorptive parts. Section \ref{CS} we present a novel calculation of the parton cross-sections of  $t\bar{t}$ production for complex CMDM and CEDM, which to our knowledge has not been reported before. In Sec. \ref{SecBounds}  a numerical simulation is presented for top quark pair production at the LHC via MadGraph5, where the effective Lagrangian of Eq. \eqref{ttgInteraction} was implemented with the help of the  FeynRules package. The results for the $t\overline{t}$ cross section as a function of the real and imaginary parts of $\hat{\mu}_t$ and $\hat{d}_t$  are then used to obtain bounds on their absorptive parts.  The possibility that kinematic distributions could be helpful to disentangle the
top quark CMDM and CEDM absorptive parts is examined in Section \ref{Distributions}. Finally, in Sec. \ref{conclusions} we present our conclusions.

\section{Remarks  on the absorptive parts of the CMDM in the SM}\label{SecComments}

\begin{figure*}[hbt!]
\begin{center}
\includegraphics[width=12cm]{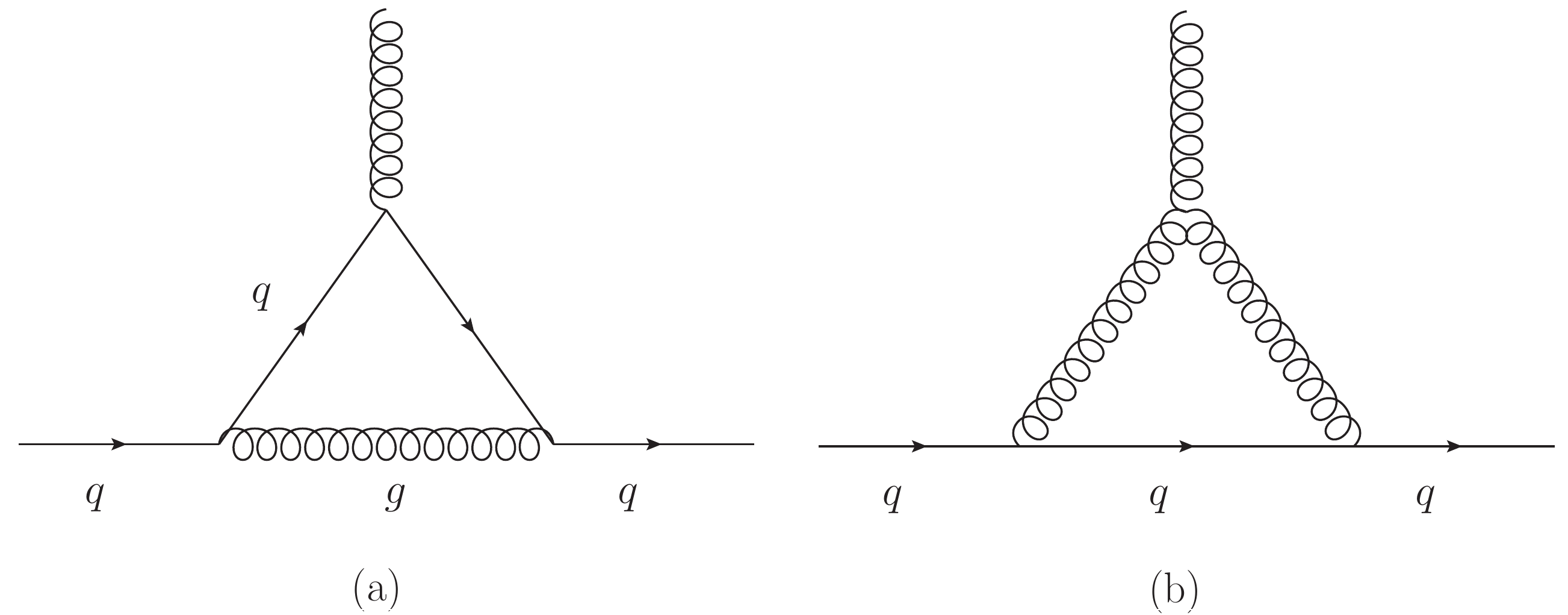}
\caption{QCD contributions to the CMDM of quarks in the SM. \label{QCD}}
\end{center}
\end{figure*}

\begin{figure*}[hbt!]
\begin{center}
\includegraphics[width=15cm]{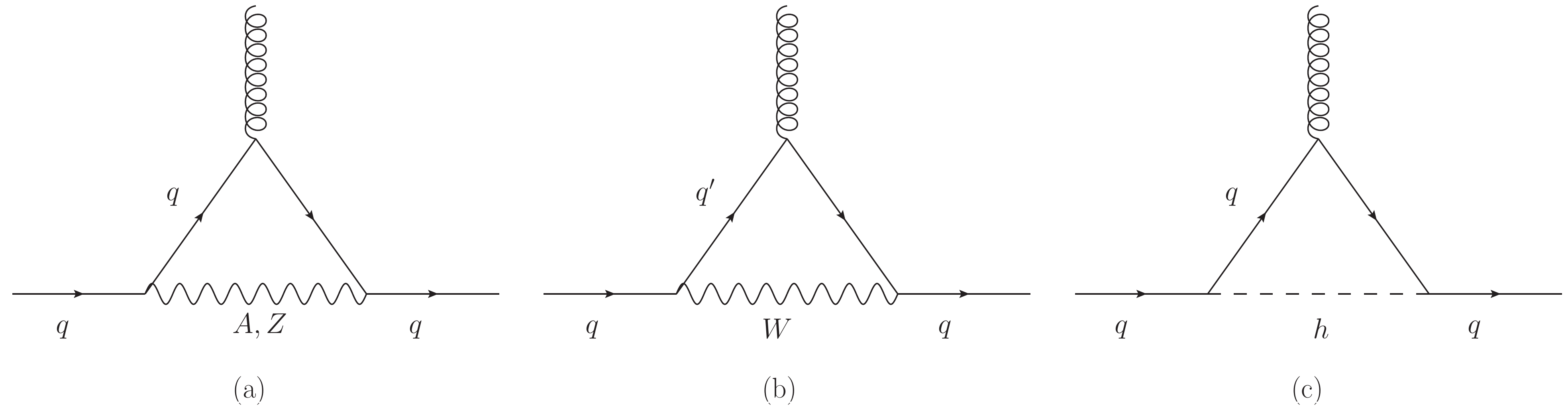}
\caption{Electroweak contributions to the CMDM of quarks in the SM. \label{EW}}
\end{center}
\end{figure*}

In the SM, the CMDM of quarks arises at the one-loop level through the Feynman diagrams of Figs. \ref{QCD} (QCD contribution) and \ref{EW} (electroweak contribution).
The off-shell CMDM  $\hat \mu_q(q^2)$ can develop an absorptive (imaginary) part when the gluon transfer four-momentum $\hat{q}=\sqrt{q^2}$ crosses the threshold $\hat{q}\ge 2m$, with $m$ the mass of the virtual particles attached to the off-shell gluon. In such region these particles are allowed to be pair produced, which is true for all energies of the external gluon in  Feynman diagram (b) of Fig. \ref{QCD}, whereas the threshold is $\hat{q}\ge 2 m_q$
for   Feynman diagrams \ref{QCD}(a),  \ref{EW}(a) and  \ref{EW}(c), and $\hat{q}\ge 2 m_{q^\prime}$ in  Feynman diagram  \ref{EW}(b).
The absorptive contributions to the CMDM can also be extracted by the Cutkosky rules \cite{Cutkosky_1960}, which yield  the same results obtained via the usual techniques for Feynman diagram calculation \cite{Zhou:2004gm}. It is also important to emphasize  that the the SM contribution to the off-shell dipole moment $\hat \mu_q(q^2)$ is finite and gauge independent for arbitrary $q^2$.


For the top quark CMDM, the contribution from the diagram (b) of Fig. \ref{EW} develops an absorptive part at $\hat{q}=2m_b$, whereas that from the diagram (b) of Fig. \ref{QCD} is complex for any $\hat{q}$ value. The remaining contributions become complex at $\hat{q}=2m_t$. Moreover,  the corresponding contributions to the top quark CEDM would also become  complex  at the same energy thresholds. It is thus interesting to obtain a bound on the absorptive part of the top quark CMDM and CEDM consistent with the CMS limits. Such a bound can be interpreted as a limit on the new physics effects inducing new contributions to the CMDM and CEDM of the top quark.
To assess the order of magnitude of the absorptive part of the CMDM at the LHC energies, we have performed  a numerical evaluation of the analytical expressions of Ref. \cite{Hernandez-Juarez:2020drn} to find the energy interval of the transfer momentum of the gluon where the real part of the top quark CMDM predicted by the SM matches  the CMS  bounds \cite{CMS:2018jcg,Sirunyan:2019eyu}. We obtain that the value ${\rm Re}\big[\hat{\mu}_t\big]=-0.024$ reported  in Ref \cite{Sirunyan:2019eyu}  corresponds to the 57 GeV$\leqslant \hat{q}\leqslant$ 59 GeV interval, where  the respective absorptive part value is ${\rm Im}\big[\hat{\mu}_t\big]\approx-0.034$. As far as the bound $-0.014<{\rm Re}\big[\hat{\mu}_t\big]<0.004$ reported in Ref. \cite{CMS:2018jcg}, it is consistent with energies above  $\hat{q}=85$ GeV, in this case, the absorptive part can be one order of magnitude smaller than in the previous one:  for values around $\hat{q}=85$ GeV, ${\rm Im}\big[\hat{\mu}_t\big]\approx-0.028$, whereas at higher energies the corresponding value is of the order of $10^{-3}$ and remains almost constant as the energy increases.  
 
\section{Contributions of  CMDM and CEDM to $\overline{t}t$ production}\label{CS}
\begin{figure*}[hbt!]
\begin{center}
\includegraphics[width=10cm]{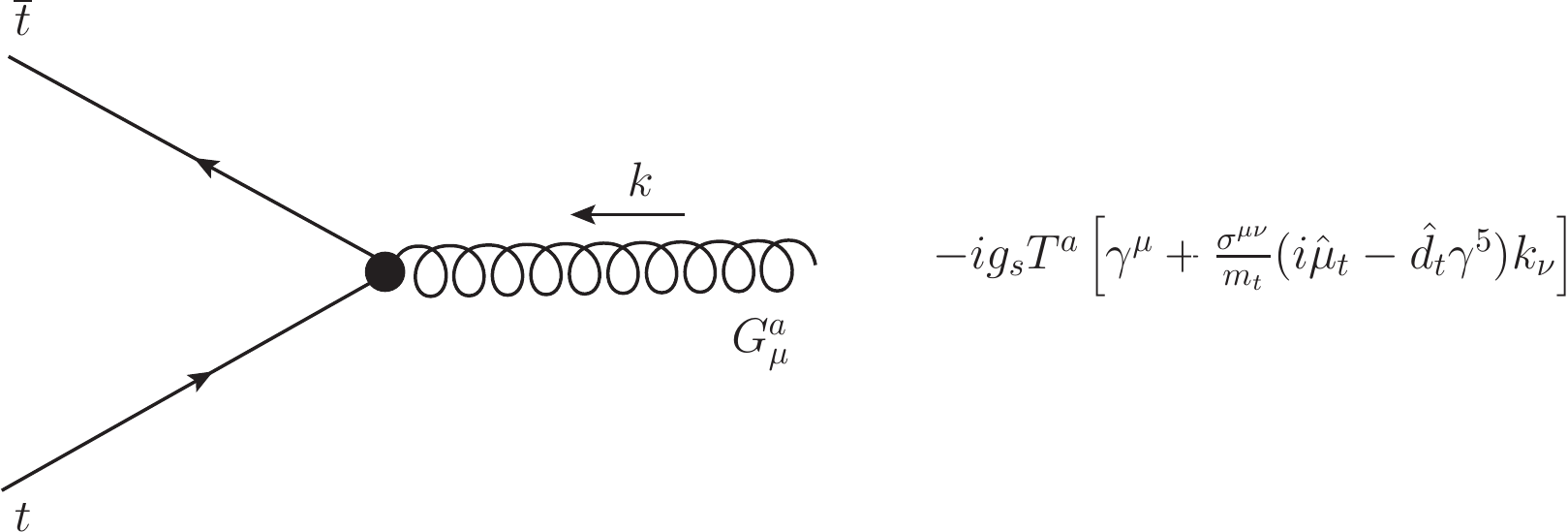}\\\includegraphics[width=10cm]{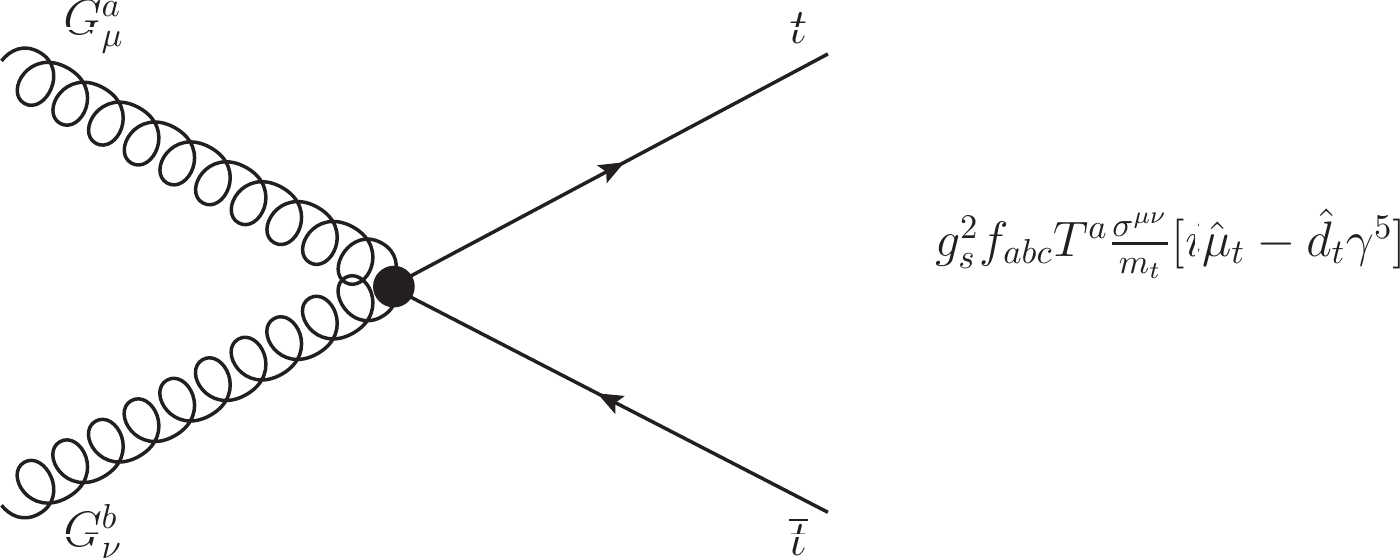}
\caption{Feynman rules for the anomalous  $\bar{t}tg$ and $\bar{t}tgg$ couplings arising from 
Lagrangian \eqref{ttgInteraction}. \label{FeynRules}}
\end{center}
\end{figure*}

Top pair production can receive contributions from  the anomalous $\overline{t}tg$ coupling \cite{Haberl:1995ek,Cheung:1995nt} of Eq. \eqref{ttgInteraction} but also from  the non-SM $\overline{t}tgg$ vertex arising from the non-abelian part of the gluon field strength tensor. The corresponding Feynman rules follow straightforwardly and are shown in Fig. \ref{FeynRules}. As already mentioned, in a strict sense,  the top quark dipole form factors are functions of the gluon transfer momentum $q^2$, with such functions being model dependent.  Furthermore, when working with  off-shell green functions one must address the problem of gauge dependence. Along this line, methods such as the pinch technique have been used in the past to remove any gauge-dependent terms by considering additional Feynman diagrams contributing to the physical process under study. However, as argued in Ref. \cite{Hernandez-Juarez:2020drn} the off-shell CMDM of the top quark is gauge-independent in the SM and the same is true for both the CMDM and CEDM in electroweak extension models. Also, as shown in Refs. \cite{Hernandez-Juarez:2020xon,Aranda:2018zis} there is little dependence of the top quark dipole form factors  on $q^2$ over a large energy interval in the SM and some BSM theories. We will thus assume that both the real and absorptive parts of the CMDM and CEDM are constant and obtain bounds on the absorptive parts from the data on top quark production at the LHC.

\begin{figure}[!htb]
\begin{center}
\subfigure{\includegraphics[width=.35\textwidth]{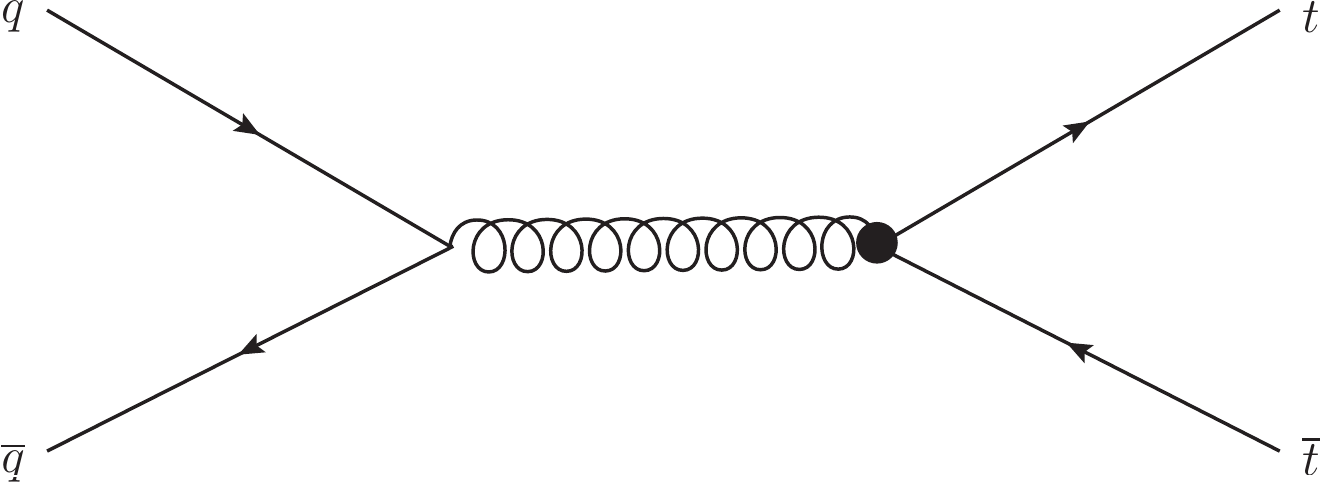}}
\subfigure{\includegraphics[width=.35\textwidth]{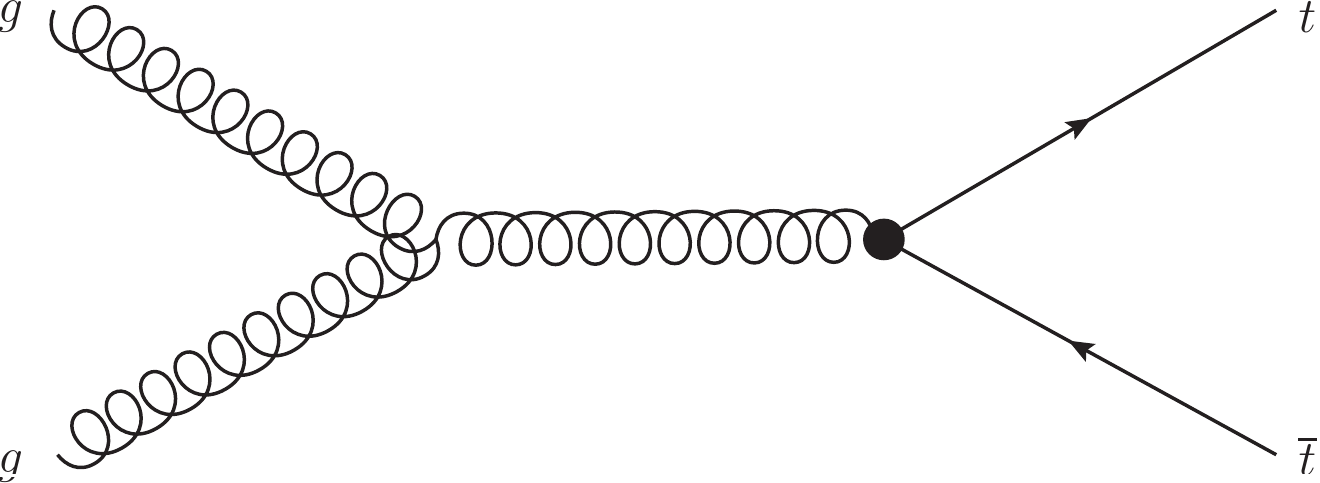}}
\subfigure{\includegraphics[width=.35\textwidth]{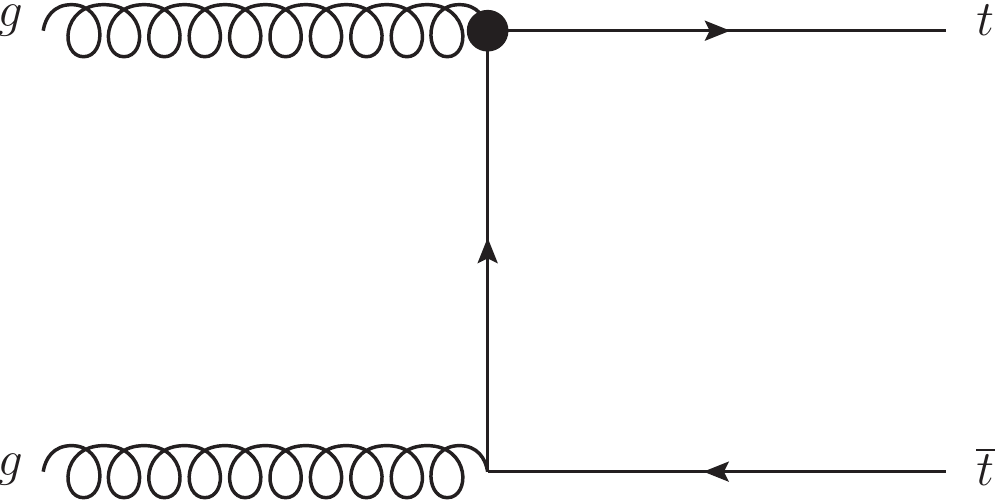}}
\subfigure{\includegraphics[width=.35\textwidth]{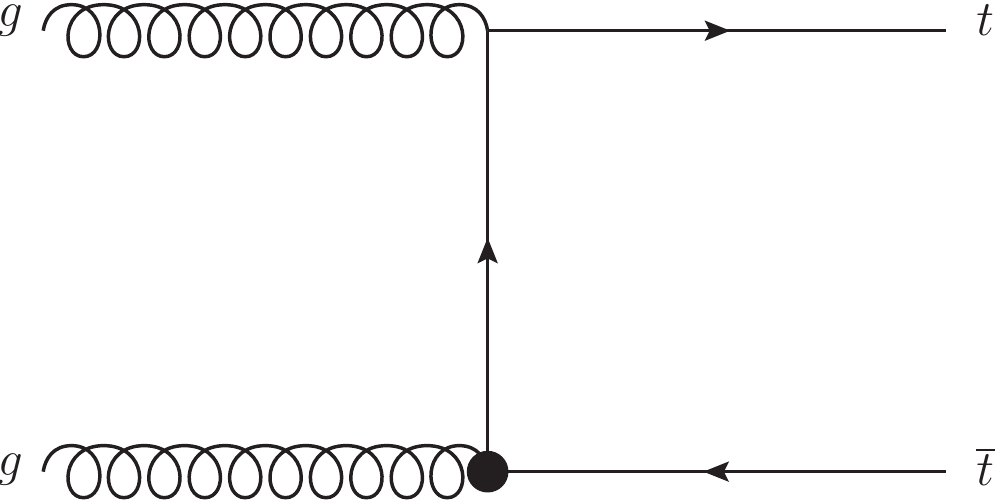}}
\subfigure{\includegraphics[width=.35\textwidth]{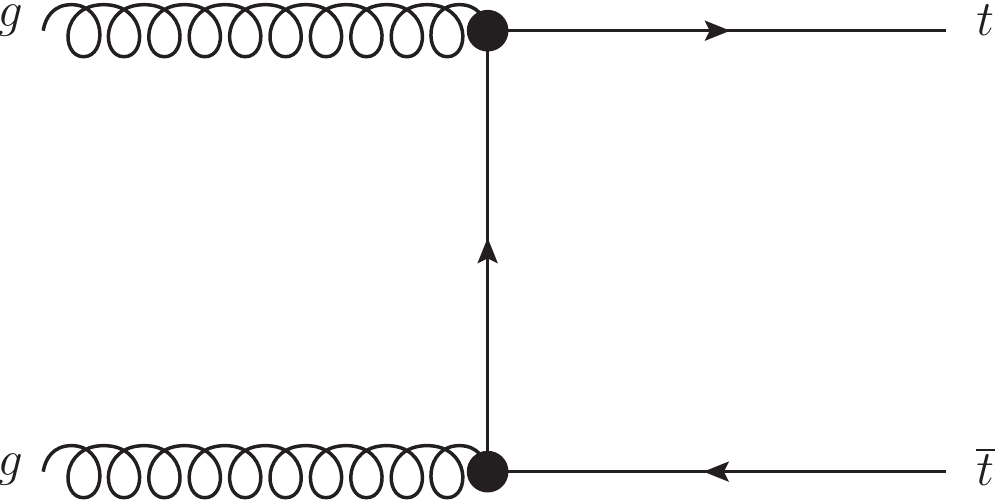}}
\subfigure{\includegraphics[width=.35\textwidth]{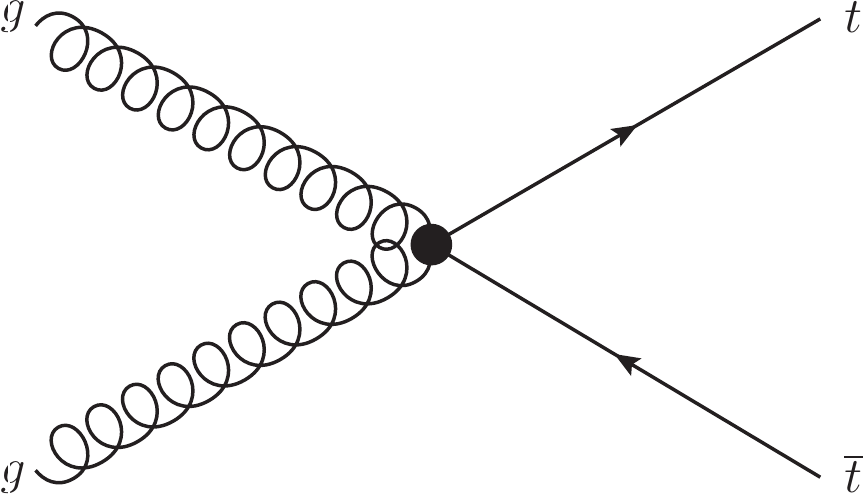}}
\caption{Feynman diagrams for the contribution to the parton cross sections $\hat{\sigma}_{q\overline{q}}$ and $\hat{\sigma}_{gg}$ at the leading order. Crossed diagrams are not shown. The large dot represents the anomalous couplings induced by the CMDM and CEDM. The SM tree-level contribution is obtained after replacing the anomalous $\bar{t}tg$ coupling by the SM one.}
\label{ttproduction}
\end{center}
\end{figure}

The most recent analyses on top quark production assume that both CMDM and CEDM are purely real \cite{CMS:2018jcg,Sirunyan:2019eyu}. In this work, we are interested in the study of the contributions of the absorptive parts of these dipole moments. Therefore we consider that   $\hat\mu_t$ and $\hat{d}_t$ are complex  and calculate  the following parton cross-sections:
 \begin{align}
    &\hat{\sigma}_{q\overline{q}}\equiv\sigma(\overline{q}q\rightarrow \overline{t}t),   \nonumber\\
    &\hat{\sigma}_{gg}\equiv  \sigma(gg \rightarrow \overline{t}t),
\end{align} 
which apart from the  SM contribution receive a new one from the Feynman diagrams of Fig. \ref{ttproduction}, where the large dot represents the anomalous CMDM and CEDM contributions. 

After some algebra we obtain the respective differential cross sections for general complex CMDM and CEDM:

\begin{align}
\label{crossqq}
\frac{d\hat{\sigma}_{q\bar{q}}}{d\hat{t}}=&\frac{\pi \alpha_s^2}{\hat{s}^2}\frac{8}{9}\Bigg[ \frac{1}{2}-v+z+2 {\rm Re}\big[\hat{\mu}_t\big]+\Big({\rm Re}\big[\hat{\mu}_t\big]^2+{\rm Im}\big[\hat{\mu}_t\big]^2-{\rm Re}\big[\hat{d}_t\big]^2-{\rm Im}\big[\hat{d}_t\big]^2 \Big)\nonumber\\
&+\Big({\rm Re}\big[\hat{\mu}_t\big]^2+{\rm Im}\big[\hat{\mu}_t\big]^2+{\rm Re}\big[\hat{d}_t\big]^2+{\rm Im}\big[\hat{d}_t\big]^2 \Big) \frac{v}{z}\Bigg],
\end{align}
and
\begin{align}
\label{crossgg}
\frac{d\hat{\sigma}_{gg}}{d\hat{t}}=&\frac{\pi \alpha_s^2}{\hat{s}^2}\frac{1}{12}\Bigg[ \Big(\frac{4}{v}-9 \Big)\Big(\frac{1}{2}-v-2z\big(1-\frac{z}{v}\big) +2\ {\rm Re}\big[\hat{\mu}_t\big] \Big)+\frac{1}{8 v z}\Big[v \Big(55\ {\rm Re}\big[\hat{d}_t\big]^2+{\rm Re}\big[\hat{\mu}_t\big]^2(55-144z)\Big) \nonumber\\
& +z\Big(4 \ {\rm Re}\big[\hat{d}_t\big]^2 +70\ {\rm Re}\big[\hat{\mu}_t\big]^2\Big)+\frac{1}{v z}\Big[-16v^3 \Big(4\Big({\rm Re}\big[\hat{\mu}_t\big]^2 {\rm Im}\big[\hat{d}_t\big]^2-4\ \ {\rm Re}\big[\hat{\mu}_t\big]\ {\rm Im}\big[\hat{\mu}_t\big]\ {\rm Re}\big[\hat{d}_t\big]\ {\rm Im}\big[\hat{d}_t\big]\nonumber\\
&+{\rm Im}\big[\hat{\mu}_t\big]^2\ {\rm Re}\big[\hat{d}_t\big]^2\Big)+9 z \Big({\rm Im}\big[\hat{\mu}_t\big]^2+{\rm Im}\big[\hat{d}_t\big]^2 \Big) \Big)+ v^2 z	\Big(-512\ {\rm Re}\big[\hat{\mu}_t\big]\ {\rm Im}\big[\hat{\mu}_t\big]\ {\rm Re}\big[\hat{d}_t\big]\ {\rm Im}\big[\hat{d}_t\big]\nonumber\\
&+ {\rm Im}\big[\hat{d}_t\big]^2\Big(16\ {\rm Re}\big[\hat{\mu}_t\big]\Big(15\ {\rm Re}\big[\hat{\mu}_t\big] +7 \Big)+288 z +63\Big)+3 \ {\rm Im}\big[\hat{\mu}_t\big]^2\Big(80\ {\rm Re}\big[\hat{d}_t\big]^2+48 z +21 \Big)\Big)\nonumber\\
&-2 v z^2\Big(92\ {\rm Im}\big[\hat{\mu}_t\big]\ {\rm Re}\big[\hat{d}_t\big]\ {\rm Im}\big[\hat{d}_t\big]+\Big(1-8\ {\rm Re}\big[\hat{d}_t\big]^2 \Big){\rm Im}\big[\hat{\mu}_t\big]^2+2\ {\rm Im}\big[\hat{d}_t\big]^2\Big(-{\rm Re}\big[\hat{\mu}_t\big]\Big(4 \ {\rm Re}\big[\hat{\mu}_t\big]+41\Big)\nonumber\\
&+72 z+17\Big) \Big)  +128\ {\rm Im}\big[\hat{d}_t\big]^2 z^3 \Big]+{\rm Re}\big[\hat{\mu}_t\big] \left( {\rm Re}\big[\hat{\mu}_t\big]^2 + {\rm Im}\big[\hat{\mu}_t\big]^2 + {\rm Re}\big[\hat{d}_t\big]^2 \right)\left(\frac{14}{z}-\frac{5}{2v} \right) \nonumber\\
 &+\left(\Big({\rm Re}\big[\hat{\mu}_t\big]^2+{\rm Im}\big[\hat{\mu}_t\big]^2\Big)^2+2\Big( {\rm Re}\big[\hat{\mu}_t\big]^2\ {\rm Re}\big[\hat{d}_t\big]^2+{\rm Im}\big[\hat{\mu}_t\big]^2\ {\rm Im}\big[\hat{d}_t\big]^2\Big)+\Big({\rm Re}\big[\hat{d}_t\big]^2+{\rm Im}\big[\hat{d}_t\big]^2 \Big)^2\right) \nonumber\\
 & \times\left(-\frac{1}{z}+\frac{1}{v}+\frac{4v}{z^2} \right)\Bigg],
\end{align}
where $\hat s$, $\hat t$ and $\hat u$ are the usual parton Mandelstam variables and we introduced the definitions
\begin{align}
z&=\frac{m_t^2}{\hat{s}},\\
 v&=\frac{1}{\hat{s}^2}(\hat{t}-m_t^2)(\hat{u}-m_t^2).
\end{align}
In the $t\overline{t}$ center of mass frame, the parameter $\hat{t}$, is related to the angle $\hat{\theta}$ between the momentum of the outgoing top quark and that of the incoming parton as
\begin{equation}\label{theta}
m_t^2-\hat{t}=\frac{\hat{s}}{2}(1-\beta \cos{\hat{\theta}}),
\end{equation}
with $\beta=\sqrt{1-4z}$.

For $\hat{\mu}_t=\hat{d}_t=0$ the above cross sections reduce to the  known SM results \cite{Gluck:1977zm} as expected. We also have verified that in the scenario with purely real CMDM and CEDM, Eq. \eqref{crossqq}  reproduces  the result reported  in Ref. \cite{Haberl:1995ek,Cheung:1995nt}. Nevertheless, in the same scenario we do not find agreement with our result for Eq. \eqref{crossgg} and the one previously reported \cite{Haberl:1995ek,Cheung:1995nt}, which  apparently is incomplete  as there is no agreement in the coefficients of  $\hat{\mu}_t^2$ and $\hat{d}^2_t$.

\section{Bounds on absorptive parts of the CMDM and CEDM of the top quark}\label{SecBounds}

We now turn to  constrain the absorptive parts ${\rm Im}\big[\hat{\mu}_t\big]$ and ${\rm Im}\big[\hat{d}_t\big]$ via the LHC data on  top quark pair  production \cite{CMS:2018jcg,Aad:2020tmz,Sirunyan:2019eyu}. We follow a similar approach to that discussed in \cite{Hayreter:2013kba} and  use a Monte Carlo simulation to obtain the theory predictions for the leading order contribution to the $\sigma(pp\to t \overline{t})$ cross section. In order to compute the corresponding contributions  from the top quark CMDM and CEDM,  we use MadGraph5 \cite{Alwall:2014hca}, where the anomalous interactions of Eq. \eqref{ttgInteraction} were implemented with the help of  FeynRules \cite{Alloul:2013bka}.

We will consider the most recent LHC results for  top quark pair production at  center-of-mass energy $\sqrt{s}=$13 TeV. Therefore we use the ATLAS cross section in the lepton plus jets channel \cite{Aad:2020tmz}
\begin{equation}\label{sigmaE}
\sigma_{\text{Exp}}(pp\to t \overline{t})=(830\pm 39) \quad\text{pb},
\end{equation}
whereas  for the theoretical SM prediction we use \cite{Zyla:2020zbs,Czakon:2008ii}
\begin{equation}\label{sigmaT}
\sigma_{\text{Theo}}(pp\to t \overline{t})=(831.8\pm43)\quad \text{pb},
\end{equation}
 wherein both cases the errors have been added in quadrature.  We will assume that the small deviation in $\sigma_{\text{Exp}}(pp\to t \overline{t})$ from the theoretical leading order SM prediction  $\sigma_{\text{Theo}}(pp\to t \overline{t})$ is due to the real and absorptive parts of the top quark CMDM and CEDM. While   90\% (85\%) of  $\sigma(pp\to t \overline{t})$ arises dominantly   from the partonic process $gg\rightarrow \overline{t}t$ at $\sqrt{s}=14$ TeV ($\sqrt{s}$=7 TeV), the contribution of $gg\overline{t}t$ vertex has not been considered in the computation of \eqref{sigmaT} as it is a non-SM interaction. However, as observed in Eqs. \eqref{crossqq} and \eqref{crossgg}, the dipole moments  induce a deviation in $\sigma(pp\to t \overline{t})$.  Thus, the top quark CMDM and CEDM may explain slight deviations from the SM prediction to  $\overline{t}t$ production.
 
 The ratio between the measured and predicted cross sections is 
\begin{equation}
\label{ratio}
\mathcal{R}=\frac{\sigma_{\text{Exp}}(pp\to t \overline{t})}{\sigma_{\text{Theo}}(pp\to t \overline{t})}=0.99\pm 0.069.
\end{equation}
Following Ref. \cite{Hayreter:2013kba}, we will interpret the error of Eq. \eqref{ratio} as a window  to  BSM effects  in top quark pair production and use it to set constraints on the absorptive parts of $\hat \mu_t$ and $\hat d_t$.  As already mentioned, in the analysis of the CMS collaboration the CMDM and CEDM of the top quark were assumed to be  real quantities. In Ref. \cite{CMS:2018jcg} all the measurements are analyzed under a linearized approximation of the anomalous couplings \cite{Bernreuther:2015yna}, which parametrizes the non-SM interactions and include the $gg\overline{t}t$ vertex, though the dipole moments are not interpreted as complex quantities and thus the absorptive parts are not considered in the differential cross section used to study BSM effects.  Also, in Ref. \cite{Sirunyan:2019eyu} the kinematic distribution used to obtain constraints on $\hat \mu_t$ and $\hat d_t$ were defined via Eqs. \eqref{crossqq} and \eqref{crossgg}, though once again it was assumed that both the top quark  CMDM and CEDM are real quantities. Thus, the effects of the absorptive parts have never been studied by the CMS collaboration, however, such consequences has  a negligible effect on their analysis as the kinematic  distributions measured at the LHC \cite{Bernreuther:2015yna,Baumgart:2012ay} are strongly constrained \cite{Fabbrichesi:2014wva,Cao:2015doa}. Below we will consider a similar approach to that followed in Refs. \cite{Hayreter:2013kba,Sirunyan:2019eyu} to obtain bounds on the absorptive parts of $\hat \mu_t$ and $\hat d_t$ by reinterpreting the LHC data. We will also show explicitly that the relevant kinematic distributions of $\overline{t}t$ production are not sensitive to the imaginary parts of the dipole moments, as already pointed out  in Refs. \cite{Bernreuther:2013aga,Bernreuther:2015yna,Baumgart:2012ay}.

To study the absorptive part of the top quark CMDM and CEDM we proceed as follows: We first set $\hat{\mu}_t=\hat{d}_t=0$ and obtain the SM cross section $\sigma_{\text{SM}}(pp\to t \overline{t})$, afterwards we
generate the new physics contribution $\sigma_{\text{NP}}(pp\to t \overline{t})$ for non-zero ${\rm Re}\big[\hat{\mu}_t\big]$, whereas all the remaining parameters are set to zero. This procedure is repeated for each one of the ${\rm Im}\big[\hat{\mu}_t\big]$, ${\rm Re}\big[\hat{d}_t\big]$ and ${\rm Im}\big[\hat{d}_t\big]$ parameters.
All our event samples for the $pp\rightarrow \overline{t}t$ cross section are generated  at $\sqrt{s}=$14 TeV.

\begin{figure*}[hbt!]

\begin{center}
\includegraphics[width=.85\textwidth]{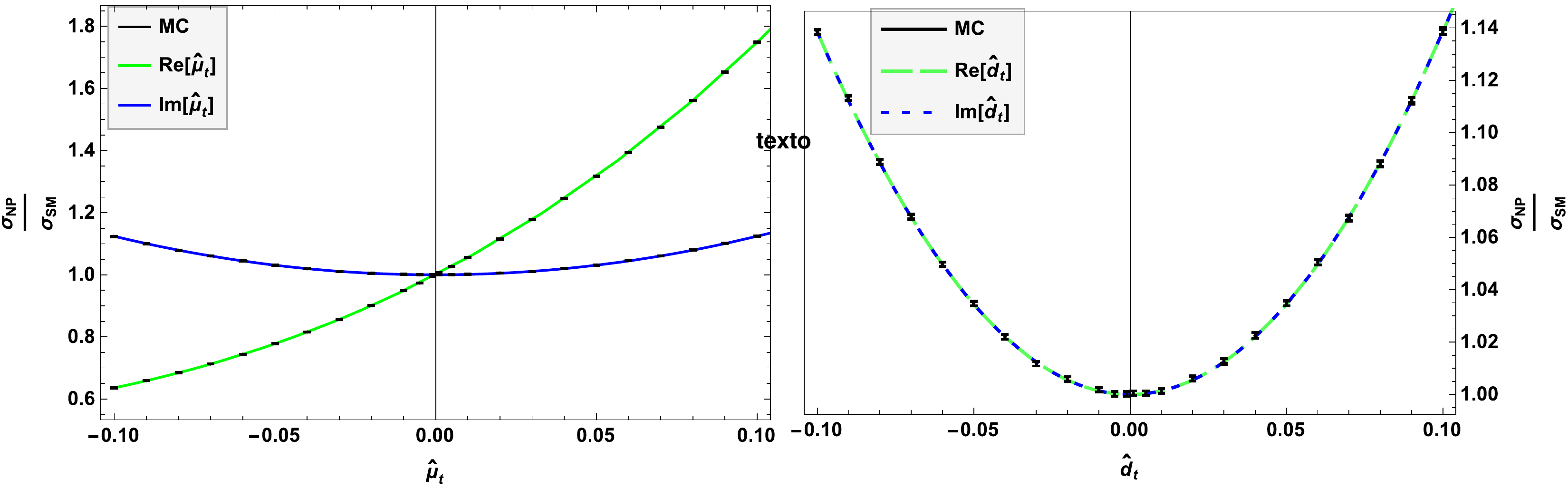}
\caption{Ratio $\mathcal{R}=\sigma_\text{NP}(pp\to \bar{t}t)/\sigma_\text{SM}(pp\to \bar{t}t)$ as a function of the real (green lines) and imaginary parts (blue lines) of the CMDM (left plot) and CEDM (right plot) of the top quark at $\sqrt{s}=14$ TeV. The bars represent the MadGraph5  estimated errors and the solid lines are the  best fit curves.\label{fit}}
\end{center}
\end{figure*}

 We show in Fig. \ref{fit} the ratio $\sigma_{\text{NP}}/\sigma_{\text{SM}}$ as a function of the real and absorptive parts of the CMDM (left plot) and CEDM (right plot), where  the  MadGraph5 estimated error is included. We plot the best fit curves.

To fit the data of Fig. \ref{fit}, we have not considered the interference terms of the real and absorptive parts of $\hat{\mu}_t$ and $\hat{d}_t$. 
Such an approach has been used in the past to study the implications of the top quark CMDM and CEDM in  $t\bar{t}$ production \cite{Hayreter:2013kba,Malekhosseini:2018fgp,Englert:2014oea}. Thus, from Eqs. \eqref{crossqq} and \eqref{crossgg} we observe that the ratio $\mathcal{R}$ is  a fourth-order polynomial of the real and imaginary parts of $\hat \mu_t$ and $\hat d_t$, though there are only even powers of the absorptive parts. The expression obtained from the fit of Fig. \ref{fit} reads
\begin{align}\label{ratioEq}
\mathcal{R}&\simeq 1 + 5.33\  {\rm Re}\big[\hat{\mu}_t\big] + 19.14\  {\rm Re}\big[\hat{\mu}_t\big]^2+21.98\  {\rm Re}\big[\hat{\mu}_t\big]^3 +5.78\  {\rm Re}\big[\hat{\mu}_t\big]^4 \nonumber\\
&+ 12.35\  {\rm Im}\big[\hat{\mu}_t\big]^2+4.38\  {\rm Im}\big[\hat{\mu}_t\big]^4+13.79\  {\rm Re}\big[\hat{d}_t\big]^2+5.58\  {\rm Re}\big[\hat{d}_t\big]^4\nonumber\\
&+ 13.78\  {\rm Im}\big[\hat{d}_t\big]^2+ 6.15\ {\rm Im}\big[\hat{d}_t\big]^4.
\end{align}
 We observe that the contributions of ${\rm Im}\big[\hat{\mu}_t\big]$, ${\rm Re}\big[\hat{d}_t\big]$ and ${\rm Im}\big[\hat{d}_t\big]$ are of the same order and  similar size, which is actually in accordance with Eq. \eqref{crossgg}.  Moreover,  $\mathcal{R}$ shows a similar dependence on the absorptive parts and  ${\rm Re}\big[\hat{d}_t\big]$, which makes it  possible to obtain bounds on ${\rm Im}\big[\hat{\mu}_t\big]$ and ${\rm Im}\big[\hat{d}_t\big]$ following Refs. \cite{Hayreter:2013kba,Sirunyan:2019eyu}, where limits on the real part of $\hat{d}_t$ are obtained using Eqs. \eqref{crossqq}, \eqref{crossgg} and \eqref{ratioEq}. We also note that the leading contribution to  $\mathcal{R}$ arises from the linear term of  ${\rm Re}\big[\hat{\mu}_t\big]$ as the other dipole terms contribute quadratically at the lowest order and are thus more suppressed. The values predicted for the off-shell top quark CMDM  are of the  order of $10^{-2}-10^{-3}$ in the SM  \cite{Hernandez-Juarez:2020drn}, whereas the typical values predicted for the CEDM  in some BSM theories are of the order of $10^{-19}-10^{-20}$ \cite{Hernandez-Juarez:2020xon}. Thus. the effects of the top quark dipole moments would hardly induce a significant deviation  to the top quark production cross section.

To analyze the effects of the absorptive parts of the top quark dipole moments we will proceed as follows. We  fix the corresponding real parts using the CMS limits \cite{CMS:2018jcg,Sirunyan:2019eyu}, which  allow us to constrain the absorptive parts ${\rm Im}\big[\hat{\mu}_t\big]$ and ${\rm Im}\big[\hat{d}_t\big]$ via Eqs. \eqref{ratioEq} and  \eqref{ratio}, with the error being attributed to the anomalous $\bar{t}tg$ contributions. In other words, we fix  ${\rm Re}\big[\hat{\mu}_t\big]$ and ${\rm Re}\big[\hat{d}_t\big]$ to their current constraints and find the allowed area of  ${\rm Im}\big[\hat{\mu}_t\big]$ and ${\rm Im}\big[\hat{d}_t\big]$ values.  With this aim we assume the following three scenarios:

\begin{itemize}
  \item Scenario I: we use the lower bounds  ${\rm Re}\big[\hat{\mu}_t\big]=-0.014$ and ${\rm Re}\big[\hat{d}_t\big]=-0.02$ reported in \cite{CMS:2018jcg}.
  \item Scenario II: we use the upper bounds  ${\rm Re}\big[\hat{\mu}_t\big]=0.004$ and ${\rm Re}\big[\hat{d}_t\big]=0.012$ reported in \cite{CMS:2018jcg}.
  \item Scenario III: we use the value ${\rm Re}\big[\hat{\mu}_t\big]=-0.024$ and the upper bound  ${\rm Re}\big[\hat{d}_t\big]=0.03$ reported in  \cite{Sirunyan:2019eyu}.
\end{itemize}
We do not consider the scenario where ${\rm Re}\big[\hat d_t\big]$ is set to its lower (negative) bound  \cite{Sirunyan:2019eyu} as it yields similar bounds to those obtained in scenario III as  $\mathcal R$ is an  even function of ${\rm Re}\big[\hat{d}_t\big]$. A similar situation arises for other possible scenarios, which  yield  bounds of a similar order of magnitude. 

\begin{figure*}[hbt!]
\begin{center}
\includegraphics[width=.55\textwidth]{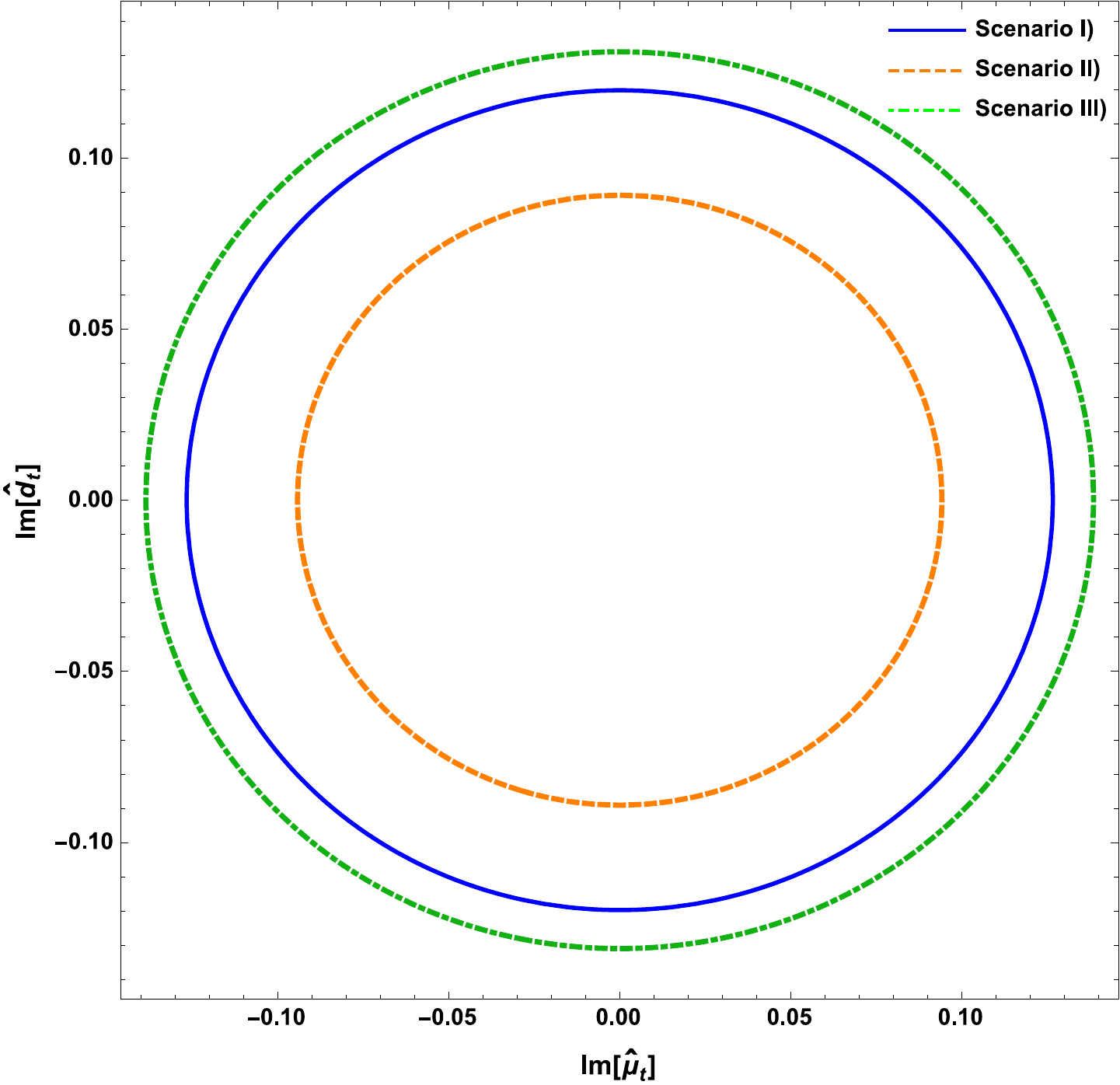}
\caption{Allowed area at the 95\% C.L. for the imaginary parts of the CMDM and CEDM of the top quark in the three scenarios  discussed in the text for the corresponding real parts. \label{bounds}}
\end{center}
\end{figure*}

 The allowed areas in the ${\rm Im}\big[\hat{\mu}_t\big]-{\rm Im}\big[\hat{d}_t\big]$ plane at the 95\% C.L. are the concentric ellipses shown in Fig. \ref{bounds} for the three scenarios discussed above. The corresponding  bounds are:  $|{\rm Im}\big[\hat{\mu}_t\big]| \lesssim 0.127$ and $|{\rm Im}\big[\hat{d}_t\big]|\lesssim 0.12$ in scenario I (blue solid lines); $|{\rm Im}\big[\hat{\mu}_t\big]| \lesssim0.139$ and $|{\rm Im}\big[\hat{d}_t\big]|\lesssim 0.133$ in scenario III (green dot-dashed lines); and $|{\rm Im}\big[\hat{\mu}_t\big]|\lesssim 0.094$ and $|{\rm Im}\big[\hat{d}_t\big]|\lesssim 0.09$  in scenario II (orange dashed line). The latter scenario yields the intersected area allowed by the three scenarios, which means that the corresponding bounds are consistent with both CMS limits. Note that in both cases  the bounds on the absorptive parts of $\hat \mu_t$ and $\hat d_t$ are quite similar,  of the order  of $10^{-1}-10^{-2}$ at the 95\% C.L. 
 
\begin{figure*}[hbt!]
\begin{center}
\includegraphics[width=.7\textwidth]{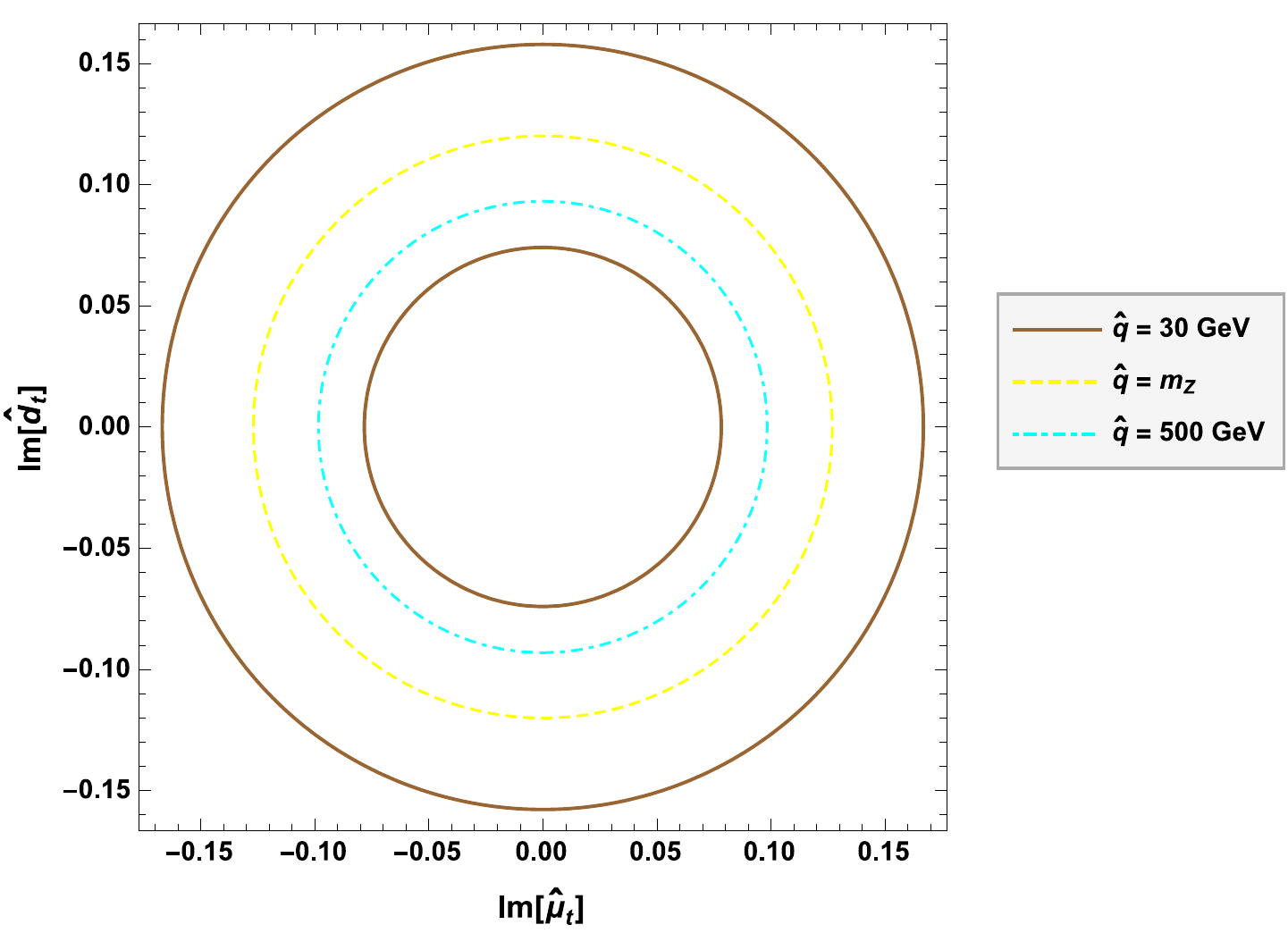}
\caption{Allowed area at the 95\% C.L. for the imaginary parts of the CMDM and CEDM of the top quark at three different energies of the gluon transfer momentum. \label{boundsQ}}
\end{center}
\end{figure*}

For completeness, we revisit the case where the gluon transfer four-momentum ($\hat{q}=\sqrt{q^2}$) dependence  is considered for  ${\rm Re}\big[\hat{\mu}_t\big]$, which is possible using the expressions  for $\hat{\mu}_t(q^2)$ reported in Ref. \cite{Hernandez-Juarez:2020drn}. The real part of the CEDM is set as   ${\rm Re}\big[\hat{d}_t\big]=0.01$ since there are not any analytic results for $\hat{d}_t(q^2)$. In Fig. \ref{boundsQ},  the allowed areas in the ${\rm Im}\big[\hat{\mu}_t\big]-{\rm Im}\big[\hat{d}_t\big]$ plane at the 95\% C.L. for three different energies of $\hat{q}$ are shown.  We note that at low energies ($\hat{q}= 30$ GeV) the bounds are slightly larger than scenario III but they are still of the same order, whereas from $\hat{q}= m_Z$ the limits are similar to those found in Fig. \ref{bounds}. Thus, the bounds  obtained by taking the real parts as constants are compatible with those where their dependence on $\hat{q}$ is considered. Especially, for energies above  the $Z$ boson mass both constraints are almost identical.  


It is worth comparing our limits with the theoretical predictions of  the SM and some BSM theories. In particular,  for a transfer momentum in the interval  30 GeV $\leqslant \hat{q}\leqslant$ 1000 GeV, the SM prediction for the absorptive part of $\hat \mu_t(q^2)$  can be as large as $10^{-2}$ \cite{Hernandez-Juarez:2020drn}, which is close to  our bounds. On the other hand, several BSM theories predict values for the absorptive part of $\hat d_t(q^2)$ of the order of  $10^{-19}$, which is far away  from  our bound.

We have also made the same analysis but including  the interference terms of Eq. \eqref{crossgg}. Nonetheless, the obtained fit is still consistent with Eq. \eqref{ratioEq}  and the  bounds are similar to those of  Fig. \ref{bounds}. Thus, the interference terms can  be neglected as their contribution is not relevant.

\section{Kinematic distributions}
\label{Distributions}

The effects of ${\rm Re}\big[\hat{\mu}_t\big]$ and ${\rm Re}\big[\hat{d}_t\big]$ on  top quark pair production have been analyzed in the past as some kinematic distributions can be sensitive to such parameters \cite{Atwood:1994vm,Cheung:1995nt,Franzosi:2015osa,Barducci:2017ddn}. On the other hand, to our knowledge the possible effects of the  absorptive parts ${\rm Im}\big[\hat{\mu}_t\big]$ and ${\rm Im}\big[\hat{d}_t\big]$ have only been explored in Ref. \cite{Bernreuther:2013aga} through the longitudinal $t$ and $\overline{t}$ polarizations. Therefore, for completeness we will examine the possibility that the differential cross sections for top quark pair production could be sensitive to the absorptive parts of the top quark CMDM and CEDM, such implications are supposed to be unobservable \cite{Bernreuther:2015yna,Baumgart:2012ay}, nonetheless it has been never shown explicitly.  To this end, we  use the CMS constraints on the real parts of the top quark dipole form factors  (${\rm Re}\big[\hat{\mu}_t\big]=-0.014$ and ${\rm Re}\big[\hat{d}_t\big] =0.01$) and analyze any possible deviation in the kinematic distributions of top quark production when both dipole moments develop an absorptive part.  We consider the following three  cases for  ${\rm Im}\big[\hat{\mu}_t\big]$ and ${\rm Im}\big[\hat{d}_t\big]$:

\renewcommand{\labelenumi}{\roman{enumi})}
\begin{enumerate}
  \item ${\rm Im}\big[\hat{\mu}_t\big]={\rm Im}\big[\hat{d}_t\big] =0.01$.
  \item ${\rm Im}\big[\hat{\mu}_t\big]={\rm Im}\big[\hat{d}_t\big] =0.05$.
  \item ${\rm Im}\big[\hat{\mu}_t\big]=0.01$ and ${\rm Im}\big[\hat{d}_t\big] =-0.01$.
\end{enumerate}
We consider such values as they are consistent with the SM prediction for the CMDM and the constraints of Sec. \ref{SecBounds}. Scenarios i) and ii) allow us to  explore the possibility that the kinematic distributions can be sensitive to small changes in the absorptive terms, whereas scenario iii) allows us to test the effect of a  flip of  sign.
For the graphical analysis, we use MADANALYSIS 5 \cite{Conte:2012fm}.

\begin{figure*}[hbt!]
\begin{center}
\subfigure[]{\includegraphics[width=.40\textwidth]{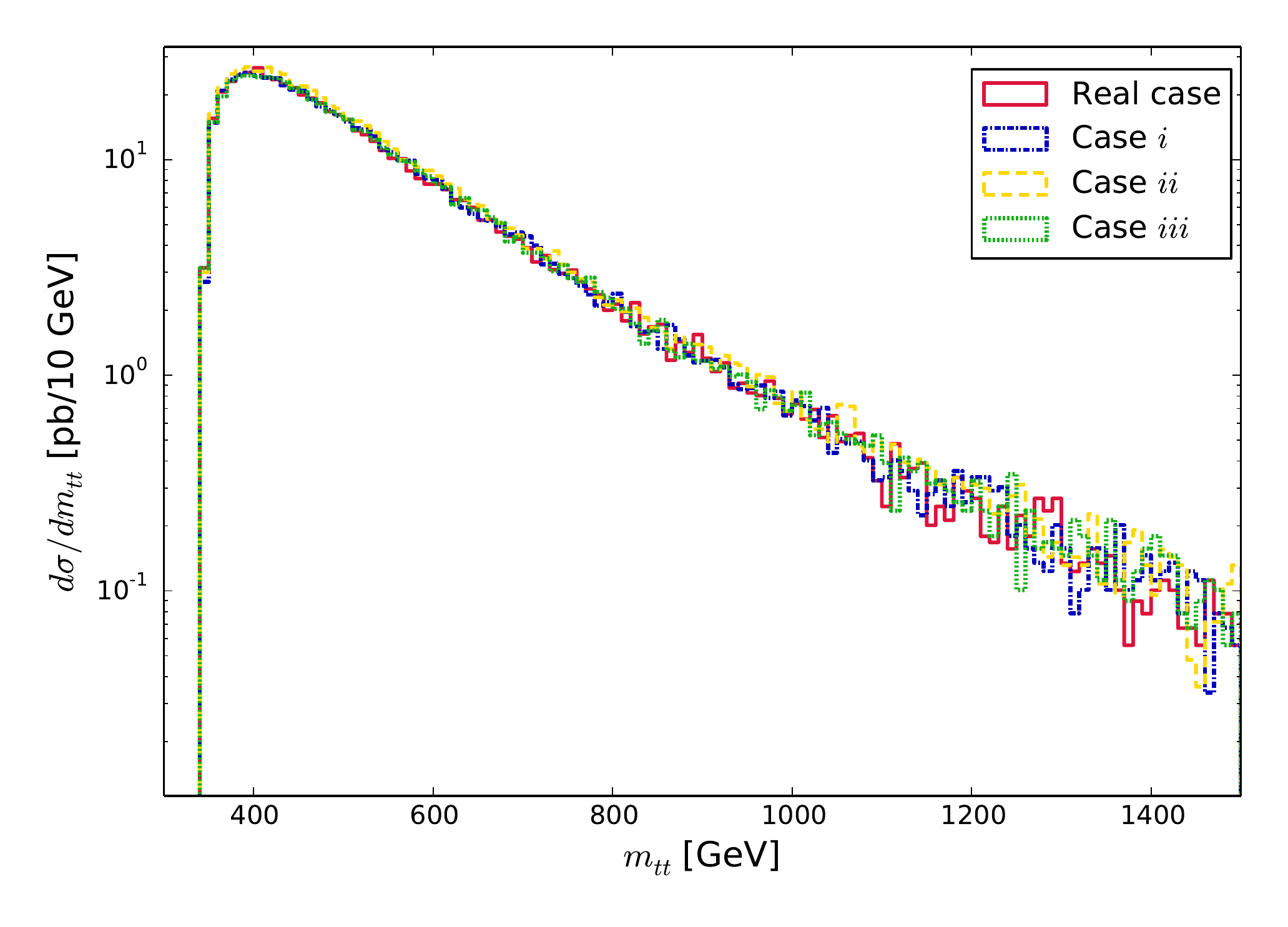}\label{mtt}}
\subfigure[]{\includegraphics[width=.40\textwidth]{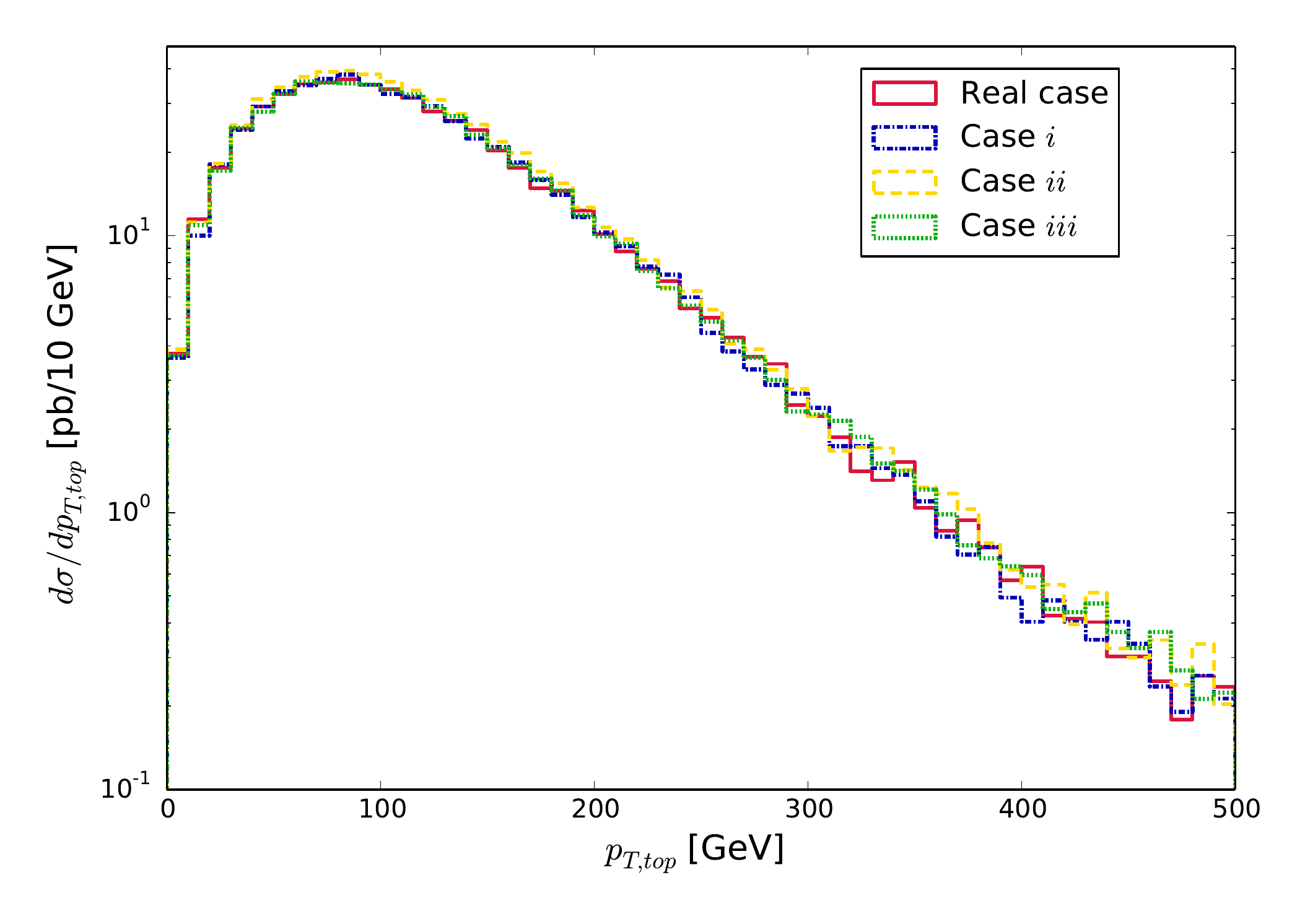}\label{pt}}
\subfigure[]{\includegraphics[width=.40\textwidth]{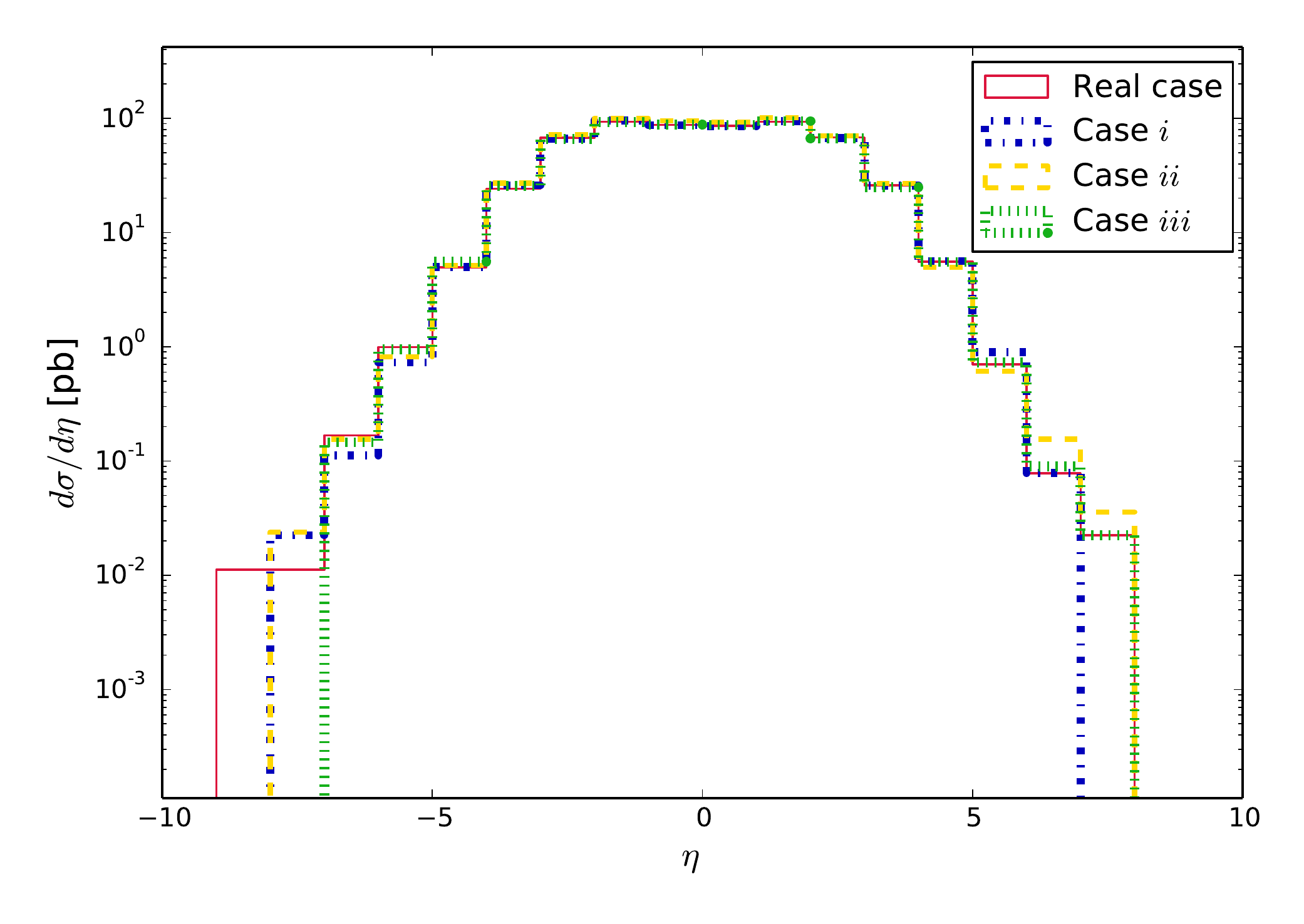}\label{eta}}
\caption{Invariant mass (a), top quark transverse momentum (b) and rapidity (c) kinematic distributions for  top quark pair  production at the LHC at $\sqrt{s}=14$ TeV in the three scenarios discussed in the text for the absorptive parts of
$\hat{\mu}_t$ and $\hat{d}_t$.}
\end{center}
\end{figure*}

In Figs. \ref{mtt} and \ref{pt} we show the kinematic distributions of the $t\overline{t}$ invariant mass  and the top quark transverse momentum in the scenarios discussed above. It is observed that there is no considerable  distinction between the kinematic distributions obtained in the general case with  complex top quark dipole form factors and those obtained in the scenario in which they are purely real. This  was  also observed in the case where  the contributions of the real part of the top quark dipole moments are compared with the SM leading order contribution \cite{Degrande:2010kt,Barducci:2017ddn}. A similar situation occurs for the kinematic distribution of the rapidity $\eta$, which is shown in  Fig. \ref{eta}.

We have also examined the  sensitivity of the forward-backward (FB) asymmetry to the CMDM and CEDM in top quark pair production at the LHC, which is possible at the leading order in some models \cite{Dorsner:2009mq,Degrande:2010kt}, whereas in the SM there  is only a significant deviation up  to next-to-leading order \cite{Dorsner:2009mq}. Unfortunately, Eqs.  \eqref{crossqq} and  \eqref{crossgg} cannot be expressed as a linear  combination of  $\cos{\theta}$ via Eq. \eqref{theta}. Thus, a deviation to the FB asymmetry at the leading order is not possible \cite{Frampton:2009rk}. However, other asymmetries could be sensitive  to the CMDM and CEDM of the top quark, as  shown in Ref.  \cite{Bernreuther:2013aga,Hayreter:2013kba}.
In summary, all the kinematic distributions studied here show no significant deviation from  leading order contribution to top quark pair production arising from the real and absorptive parts of  the top quark dipole form factors.

\section{Conclusions}\label{conclusions}
The off-shell CMDM and CEDM of quarks  have become a topic of interest recently \cite{Hernandez-Juarez:2020drn,Aranda:2020tox}. However,  the study of their absorptive (imaginary) parts remains almost unexplored.  In this work, we  have obtained bounds on the new physics contributions to the absorptive parts of the off-shell top quark CMDM and CEDM via the experimental data of top quark pair production at the LHC, which to our knowledge are the first limits of this kind. We present explicit expressions for the corresponding differential parton cross-sections considering  complex CMDM and CEDM, which have also been calculated for the first time. We point out that there is a disagreement between our result for the $gg\to t\bar{t}$ differential cross section and the expression previously reported in the scenario where only the real part of the top quark dipole form factors are considered \cite{Haberl:1995ek,Cheung:1995nt}. Our bounds for the absorptive parts were obtained using the  most recent data for the top quark CMDM and CEDM reported by the CMS collaboration \cite{CMS:2018jcg,Sirunyan:2019eyu}. It was found  that the upper bound on the absorptive parts of both dipole moments are of  order  $10^{-1}-10^{-2}$. In particular, values of  order  $10^{-2}$ are consistent with all the CMS results.  We also note that our bound  on ${\rm Im}\big[\hat{\mu}_t\big]$ is consistent with the SM prediction for the absorptive part of $\hat\mu_t$, which is of  order  $10^{-2}-10^{-3}$ \cite{Hernandez-Juarez:2020drn,Aranda:2020tox}. On the other hand, in some BSM theories the absorptive part of the CEDM is of the order of $10^{-19}$ \cite{Hernandez-Juarez:2020xon}, which seems well beyond the experimental reach. Our  limits could be useful to constrain the parameter space of BSM theories  and  are consistent with the case where the gluon transfer four-momentum ($\hat{q}$) dependence is considered for the real part of the top quark CMDM.

We also explored the possibility that several kinematic distributions for top quark pair production at the LHC can be sensitive to the absorptive parts of the CMDM and CEDM, but we find that there are no significant deviation from the scenario where the CMDM and CEDM are purely real. In fact, even in the case of real CMDM and CEDM, there is no significant deviation from the leading order SM contribution as discussed previously  \cite{Degrande:2010kt,Barducci:2017ddn}.

\section{acknowledgements}
We acknowledge support from Consejo Nacional de Ciencia y Tecnolog\'ia and Sistema Nacional de Investigadores. Partial support from Vicerrector\'ia de Investigaci\'on y Estudios de Posgrado de la Ben\'emerita Universidad Aut\'onoma de Puebla is also acknowledged. We also thank to M. A. Arroyo-Ure\~na for discussions  to implement our model in MadGraph5.

\section*{Data Availability Statement}
The datasets generated during and/or analysed during the current study are available from the corresponding author on reasonable request.

\bibliographystyle{unsrt}
\bibliography{articleN}

\end{document}